\documentclass[review,number,sort&compress,3p,times]{elsarticle}	
 \usepackage{lineno}
 \usepackage{color}

\linenumbers 
\usepackage{amssymb}
\usepackage{subfigure}
\begin{document}
\begin{frontmatter}
\title{Structural characterization of ``as-deposited'' cesium iodide films studied by X-ray diffraction and transmission electron microscopy techniques}
\author{Triloki, P. Garg, R. Rai}
\author {B. K.~Singh\corref{cor}}
\cortext[cor]{Corresponding author}
\ead{bksingh@bhu.ac.in}
\address{High Energy Physics laboratory, Physics Department,
  Banaras Hindu University,
  Varanasi 221005 India}

\begin{abstract}
  In the present work, cesium iodide (CsI) thin films of different thickness have been prepared by thermal evaporation technique. The crystallite size and grain size of these films are compared by using X-ray diffraction (XRD) profile analysis as well as by transmission electron microscopy (TEM) counting, respectively. These two methods provide less deviation between crystallite size and grain size in the case of thin CsI films of 4 nm, but there is comparatively large difference in case of thicker CsI films (20 nm, 100 nm and 500 nm). It indicates that dislocations are arranged in a configuration which causes small orientational difference between two adjacent coherent regions.  The crystallite size obtained from XRD  corresponds to the size of the coherent scattering region, whereas in TEM micrograph, single grain may correspond to many such coherent scattering regions. Other physical parameters such as strain, stress and deformation energy density are also estimated precisely for the prominent XRD peaks of thicker CsI films in the range $2\theta =  20^{0}-80^{0}$ by using a modified Williamson-Hall (W-H) analysis assuming uniform deformation model (UDM), uniform deformation stress model (UDSM) and uniform deformation energy density model (UDEDM). 
  
\end{abstract}

\end{frontmatter}

\section{Introduction}
Alkali halide materials are of technological importance due to their excellent electron-emitting properties in the ultraviolet (UV), vacuum ultraviolet (VUV),   extreme ultraviolet (EUV) and X-ray energy ranges. These materials are currently employed in vacuum and gas-based photon detectors~\cite{C.Lu,D. Morrman}, detection of scintillation light~\cite{Daisuke}, medical imaging~\cite{Wei}, positron emission tomography~\cite{F. Garibaldi} as well as a protective layer in visible-sensitive photon detectors~\cite{A. Breskin1996}.  Among alkali-halide photocathodes, CsI is the best choice, owing to its high quantum efficiency (QE) in the VUV wavelength range ~\cite{A. Breskin1997, A. Breskin1995}. CsI films are also used to enhance the field emission (FE) sources which have potential applications including display devices ~\cite{V. Vlashos}, X-ray tubes~\cite{Toru}, charged particle accelerators~\cite{A. Jhingan} and high power microwave devices ~\cite{R.J. Umstattd}. Shiffler et al~\cite{D.A. Shiffler2002} has reported  a reduction in outgassing and improved emission uniformity after CsI coatings on carbon fibers. Even two orders of magnitude reduction in turn-on voltage was successfully achieved by means of CsI coating on carbon fiber-based FE devices by the same group ~\cite{D.A. Shiffler2008}.  Due to the importance of CsI photocathodes,  several thin film preparation methods, such as thermal evaporation~\cite{V. Dangendorf, J. Seguinot}, ion beam sputtering~\cite{M.A. Nitti2}, e-gun evaporation~\cite{P. Maier-Komor}, spray pyrolysis~\cite{S.O. Klimonsky}, pulsed laser deposition~\cite{S.B. Fairchild} are used to study the various physical and chemical properties of CsI. However, it has been observed that the thermal evaporation is the best choice forming a stoichiometric Cs:I ratio ~\cite{B. K. Singh} as well as the highest absolute quantum efficiency (QE) compared to other preparation techniques~\cite{M.A. Nitti2, P. Maier-Komor, S.O. Klimonsky, S.B. Fairchild}. Even with its enormous applications in a variety of fields discussed above, very few of the earlier studies in this field deal with characterization of CsI film structure~\cite{A.S. Tremsin,MA Nitti,Coluzza, J.Almeida, H. Hoedlmoser,triloki}. X-ray diffraction (XRD) and transmission electron microscopy (TEM) are the two important techniques which are commonly used for the structural characterization.

 The XRD Peak profile analysis endeavors to characterize microstructural features of the sample from the shape and breadth of Bragg's diffraction peaks, which arise due to finite crystallite size and microstrain. As broadening due to finite crystallite size and microstrain occurs together, various analytical method, such as Variance method~\cite{florentino}, Warren-Averbach method ~\cite{Warren} and Williamson-Hall analysis~\cite{G. K.Williamson}, have been adopted to separate both effects. Among all available methods, Williamson-Hall is a simplified approach to deconvolute strain and finite size induced broadening by plotting the total breadths of the reciprocal lattice point against their distance from origin ~\cite{langford}. On the contrary, Variance and Warren-Averbach methods are more complex to analyze and their application is restricted to materials having high symmetry or which exhibit a high degree of prefered orientation. Therefore, in present manuscript, we emphasized on  W-H method to study the variation of crystallite size with thickness of the films and to separate the strain and finite size induced braodening.

  In Williamson-Hall method, broadening in Bragg's peak is assumed to be the sum of peak broadening due to finite crystallite size and induced strain. If strain is assumed to be uniform in all crystallographic directions then W-H model turns to uniform deformation model (UDM). In UDM, all the material properties are independent of the direction along which they are measured. Further, in uniform deformation stress model (UDSM) the strain is assumed to have a linear proportionality with stress according to Hook's law.  UDSM is an approximation which is valid only for the small strain present in the crystal. Another model, uniform deformation energy density model (UDEDM) is used to determine the energy density of a crystal. In this approach the crystals are assumed to have a homogeneous and isotropic distribution. However, this assumption does not hold good and constants of proportionality associated with stress-strain relation are no longer independent when stress energy density is considered.

The present paper accounts for the surface characterization of as-deposited CsI thin films of different thickness prepared by thermal evaporation technique.  The characterization of crystalline materials mainly comprises the description of grain size and internal stress or strain due to various lattice defects~\cite{Tamas ungar}. Usually the size obtained by XRD corresponds to the average of the smallest undistorted region in the material whereas TEM counting is related to regions separated by continuous boundaries in the TEM micrograph. To distinguish the two sizes, we will use terms as crystallite size for XRD and grain size for TEM results. A comparative evaluation of the mean grain size of as-deposited CsI thin films obtained from direct TEM measurement, as well as the the crystallite size obtained from Williamson-Hall method using XRD measurement is studied. In addition, the strain associated with the as-deposited CsI films due to lattice deformation is estimated by a modified form of Williamson-Hall approach namely uniform deformation model (UDM). The other modified models such as UDSM and UDEDM are also used to provide an idea of the stress as well as the uniform deformation energy density.

\section{Experimental Details}

The experimental setup for CsI consists of a high vacuum evaporation chamber which includes an oil-free Pfeiffer-made pumping unit equipped with a turbo-molecular pump having a pumping speed of 510 liter/second for $N_{2}$ and a diaphragm pump. Base pressure of this vacuum chamber is of the order of $3\times 10^{-7}$ Torr. Small pieces of CsI crystal were placed in a tantalum boat inside the chamber and carefully heated to allow out-gassing from the surface of the crystal, if any, under a shutter. After proper out-gassing and melting of CsI crystals, thin films of different thickness were deposited on polished aluminum (Al) substrates and formvar coated copper (Cu) grids. Before deposition, typical composition of different residual gases including water vapor inside the chamber were monitored through a residual gas analyzer (SRS RGA 300 unit) as shown in Figure 1. It has been confirmed that the amount of water vapor inside the vacuum chamber was under controlled manner.
\begin{figure}[!ht]
  \begin{center}
    \includegraphics[scale=0.75]{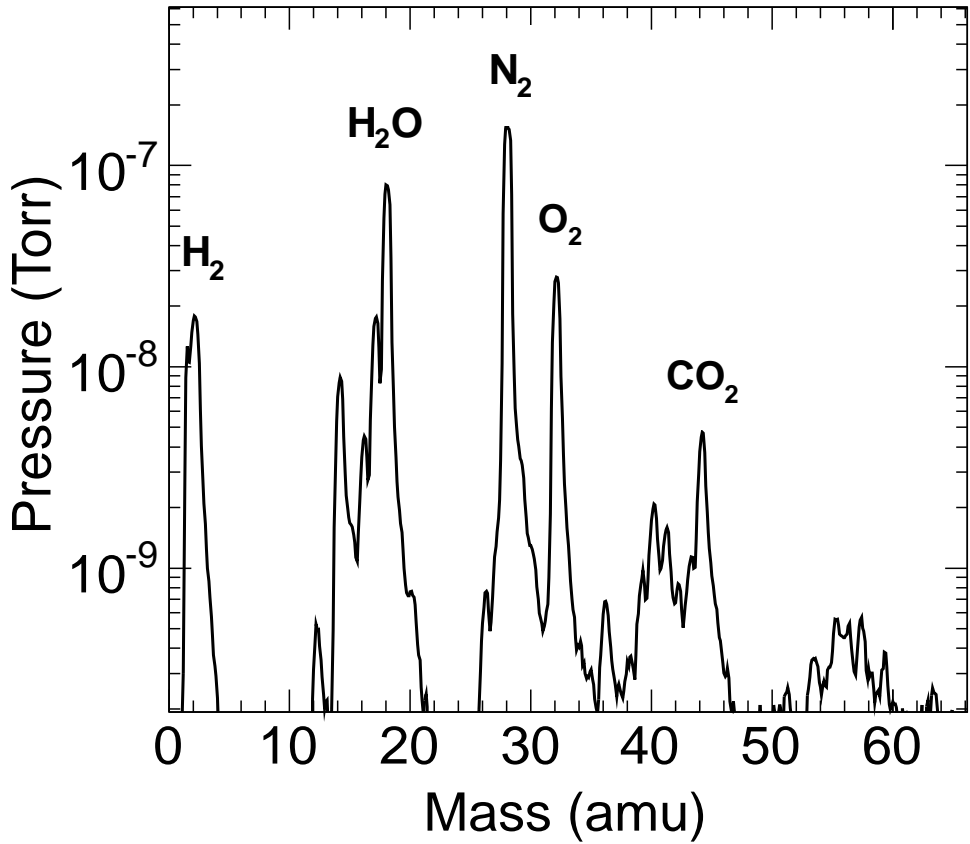}
    \caption{Residual gas composition inside the vacuum chamber.}
    \label{fig1}
  \end{center}
\end{figure}
During the film deposition, the rate of evaporation was about 1-2 nm per second and the boat and substrate were kept at a distance of about 20 cm. The thickness of the film was controlled by a quartz crystal thickness monitor (Sycon STM100).

After film deposition, the vacuum chamber was purged with nitrogen $(N_{2})$ gas in order to avoid the effect of humidity on the prepared CsI samples. Immediately after the chamber opening under constant flow of $N_{2}$, as-deposited CsI thin films were extracted and placed in a vacuum desiccator.  Further, CsI films deposited on formvar coated copper grid were used for TEM measurement while those deposited on Al substrate for XRD measurement.

The structural measurements were performed by X-ray diffraction (XRD) technique in the Bragg-Brentano parafocussing geometry using PANalytical XÕPert PRO XRD system. The incident beam optics consists of a $CuK_ {\alpha}$ radiation source ($\lambda=1.5406\AA$) and a nickel (Ni) filter. XRD measurements have been performed in continuous scan mode in the range $2\theta =  20^{0}-80^{0}$. The diffracted beam optics consists of a 0.04 rad solar slit and a scintillator detector. Similarly, transmission electron microscopy (TEM) measurements were done by means of FEI Tecnai $20G^{2}$ operating at 200 KV voltage for the examination of structure and grain size of CsI films. 

\section{Results and discussion}

\subsection{Crystallite size and strain by XRD analysis}

XRD patterns of cesium iodide thin films with different thickness prepared by thermal evaporation technique are shown in Figure 2.
\begin{figure}[!ht]
  \begin{center}
    \includegraphics[scale=0.75]{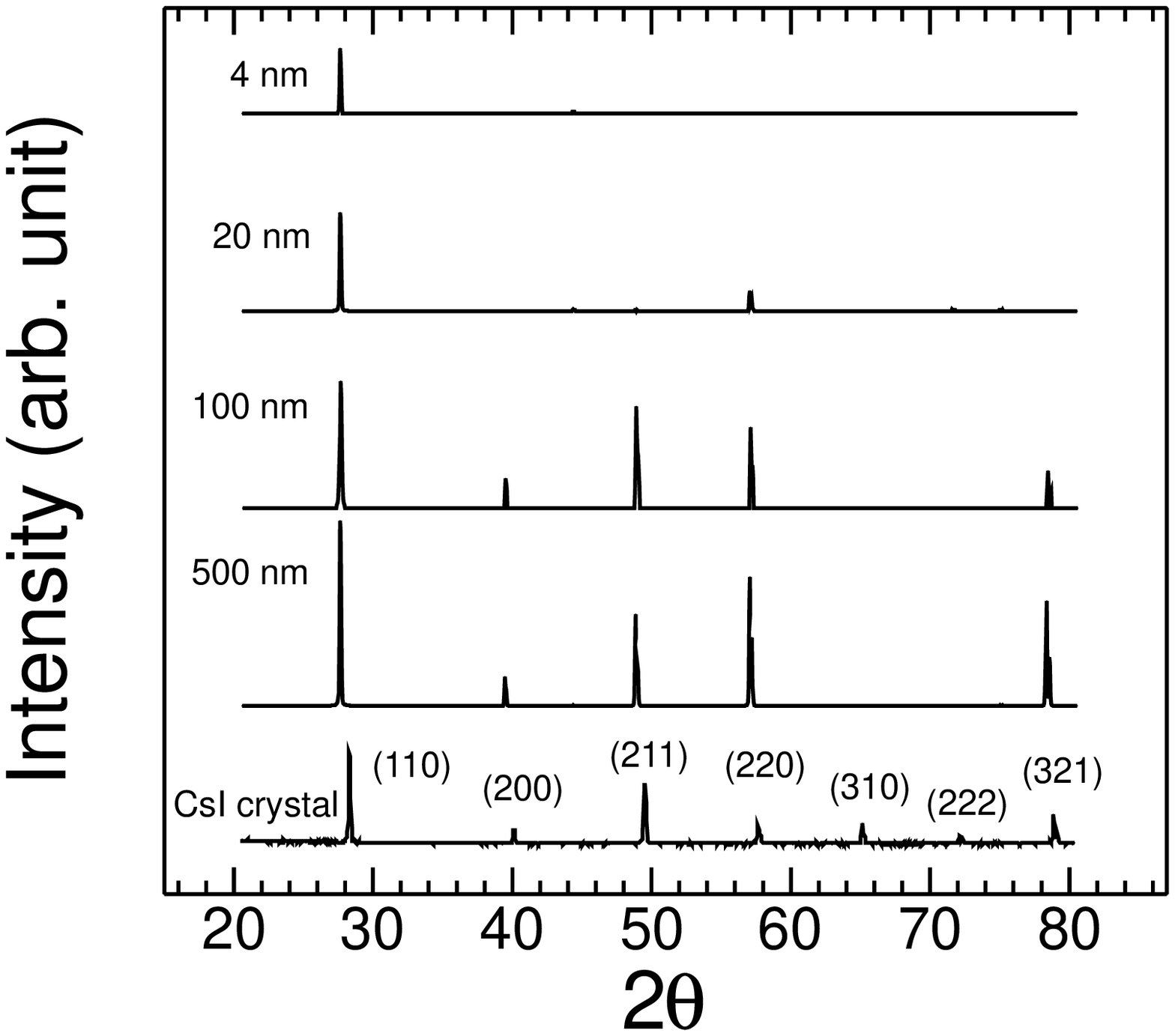}
    \caption{X-ray diffraction pattern of CsI thin films of different thickness, deposited on aluminum substrate and of CsI crystal.}
    \label{comparision_thickness.eps}
  \end{center}
\end{figure}
No extra diffraction peaks corresponding to Cs, $Cs_{2}O$, $CsIO_{3}$ or other CsI phases are detected indicating that pure CsI is of polycrystalline, stoichiometric nature. Further, the XRD result of raw CsI crystal used for thermal evaporation is shown for comparison. The XRD scan exhibits a number of intense and sharp peaks which are assigned to the indicated Bragg reflections from CsI crystal. We may observe that the lattice plane corresponding to the preferred peaks for CsI crystal are: (110), (200), (211), (220), (310), (222) and (321). In the case of 4 nm Òas-depositedÓ CsI thin films, we observe the peak of (110)  lattice plane only. In case of 20 nm Òas-depositedÓ film, we observe the lattice planes of (110) and (220) only. However, for thicker Òas-depositedÓ CsI films (100 nm and 500 nm), most intense peaks of (110) followed by (200), (211), (220) and (321) can be clearly observed.  As peak (321) is contaminated by (311) peak of aluminum substrate, it is excluded from the present analysis. These peaks match with the peak positions listed for cesium iodide in ASTM card No. 060311, confirming the films to be of CsI.   The value of full width at half maximum (FWHM) and $2\theta$ corresponding to the most intense (110) peak for various thickness of thin CsI films are shown in Table 1. Using XRD profile  (shown in Figure 2), lattice parameters of CsI crystal as well as CsI thin films are calculated. The lattice constant (a) for all thicknesses of CsI film is obtained as $4.666\AA$, however lattice constant for CsI crystal is about $4.566\AA$.

\begin{table}[ht]
  \begin{center}
    \caption[]{The FWHM and $2\theta$ corresponding to different thicknesses of CsI film for most intense (110) peak.}
    %%\begin{fntable}[0.6\columnwidth]
    \begin{tabular}{|c|c|c|c|}
      \hline
      Thickness & FWHM & $2\theta$ \\\hline
      500 nm & 0.1476     & 27.0594 \\
      100 nm &0.1476      & 27.0617 \\
      20 nm & 0.1476     &27.0596 \\
      4 nm & 0.1968 & 27.0797 \\
      \hline
    \end{tabular}
    %%\end{fntable}
    %%\caption[]{The values of FWHM and $2\Theta$ corresponding to diffrerent thicknesses of thin film CsI.}
    \label{tab:fntable} 
  \end{center} 
\end{table}

\begin{figure}[!ht]
  \begin{center}
    \includegraphics[scale=0.75]{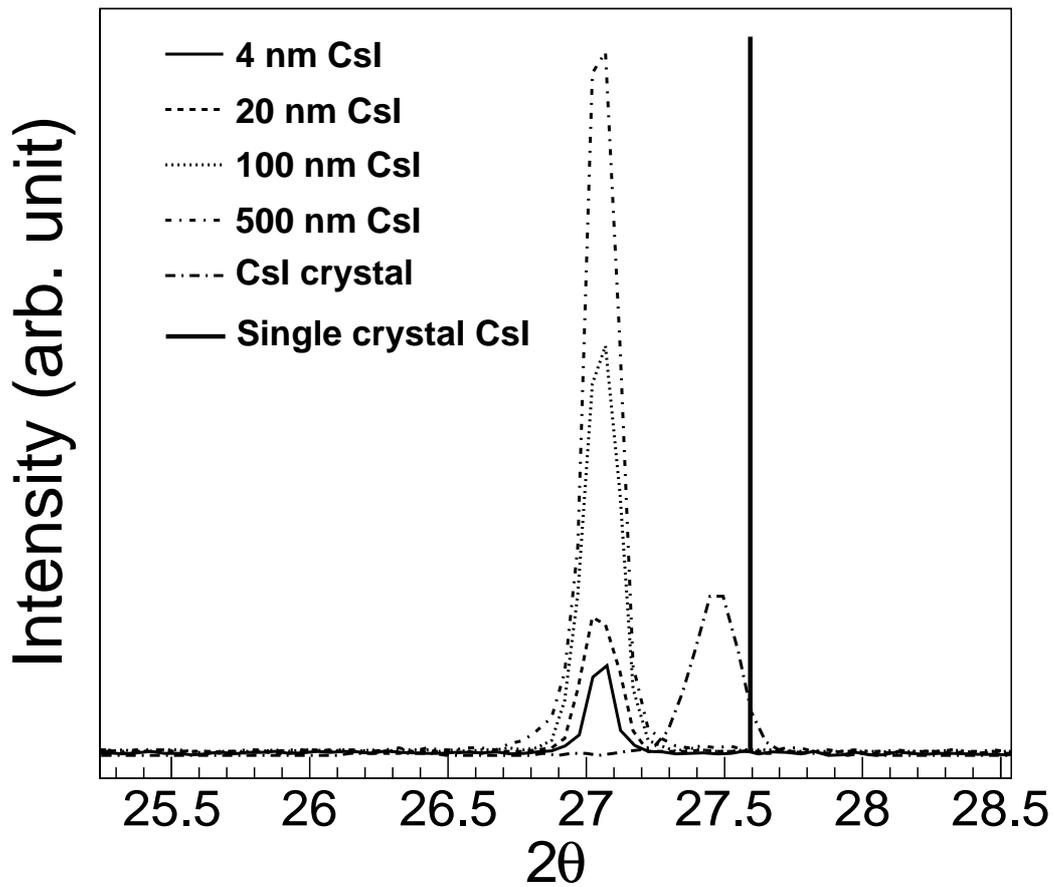}
    \caption{Shifts in the (110) peaks of the X-ray diffraction pattern as   compared to single crystal shown with sharp solid line.}
    \label{xrd_peak_shift}
  \end{center}
\end{figure}

The average crystallite size is calculated by using {the Debye-Scherrer}'s equation ~\cite{P. Scherrer} as follows:

\begin{equation}
  D= \frac{k \lambda }{ \beta_{hkl} \cos\theta} 
\end{equation}

where D is the volume weighted crystallite size, k is the shape factor (0.89), $\lambda$ is the wavelength of $CuK_{\alpha}$ radiation, $\beta_{hkl}$ is full width at half maximum (FWHM) of the particular peak and $\theta$ is the {Bragg}'s angle. From the calculations, the average crystallite size of CsI thin films are obtained as 41 nm and 55 nm for 4 nm and 20 nm thin films respectively, while for 100 nm and 500 nm thick CsI films it is obtained to be 54.74 nm. The crystallite size obtained by us is in good agreement with the reported crystallite size of 45 nm for 100 nm CsI thin films by Nitti et al ~\cite{M.A. Nitti} using same thermal evaporation technique. Klimonsky et al ~\cite{S.O. Klimonsky} have also reported the crystallite size of about 45-50 nm for different CsI samples prepared by spray pyrolysis technique. However, for same thickness of 100 nm film deposited by means of ion beam sputtering and ion beam assisted sputtering techniques, Nitti et al ~\cite{M.A. Nitti2} have reported the increased crystallite size of about 334 and 288 nm respectively. 

 Further, the crystallite size depends on the broadening of the diffracted peak and Williamson-Hall approach~\cite{G. K.Williamson} allows us to find two different reason for it: One is the finite crystallite size, which varies as $1/\cos\theta$ (see equation 1), the other is the induced strain ($\epsilon$), which is given by Wilson formula ( $\beta_{hkl} = 4 \epsilon\tan\theta$)~\cite{A.R. Stokes}. Therefore, XRD profile can be used to determine residual stress and strain in the sample and the apparent shift in diffraction patterns from their corresponding crystal data indicates a uniform stress originated in the film due to the thermal evaporation  ~\cite{G.C. Budakoti, P. Arun}. A shift in the peak position is also observed in our CsI films as shown in Figure 3 for (110) plane in comparison with the peaks observed in XRD scan of CsI crystals. It indicates that microstrain has developed in the prepared thin films. In our case, CsI (110)  peaks are shifted towards lower angles of $\theta$ as compared to the crystal data (2$\theta=27.592^{0}$) from ASTM card No. 060311 as shown in Figure 3. These stresses acting in the film arise due to the various methods of film preparation and can cause some effects on the properties of the materials, in particular photoemissive properties are affected by the method of film preparation as shown and discussed in reference~\cite{M.A. Nitti2}.

In Williamson-Hall approach the line broadening due to finite size of coherent scattering region and the internal stress in the prepared films are considered.  The finite size is taken care by Scherrer's equation and the stress by Wilson formula in Williamson-Hall equation as follows~\cite{G. K.Williamson}.

\begin{equation}
  \beta_{hkl}\cos\theta = \frac{k\lambda}{D} + 4 \epsilon\sin\theta
\end{equation}

where $\epsilon$ is the strain, which is usually assumed to be proportional to the square root of the density of dislocations, $\beta cos\theta/\lambda$ is the total integral breadth in reciprocal space and $2sin\theta/\lambda$ is the distance of reciprocal point from the origin.  Figure 4(a) and 4(d) shows the measured values of $\beta_{hkl} \cos\theta$ as a function of $4\sin\theta$ for 500 nm and 100 nm CsI films. One can estimate the strain from the slope of the fitted line and crystallite size (D) from its intersection with the ordinate. Equation (2) corresponds to uniform deformation model, which consider the isotropic nature of crystal. In Table 2, it is shown that the strain as well as the estimated crystallite size obtained for 100 nm is more than the 500 nm film (see Table 2 for details). It indicates that by increasing the thickness of CsI film strain and crystallite size decrease. 

\begin{figure*}[!ht]
  \begin{center}
    \includegraphics[scale=0.6]{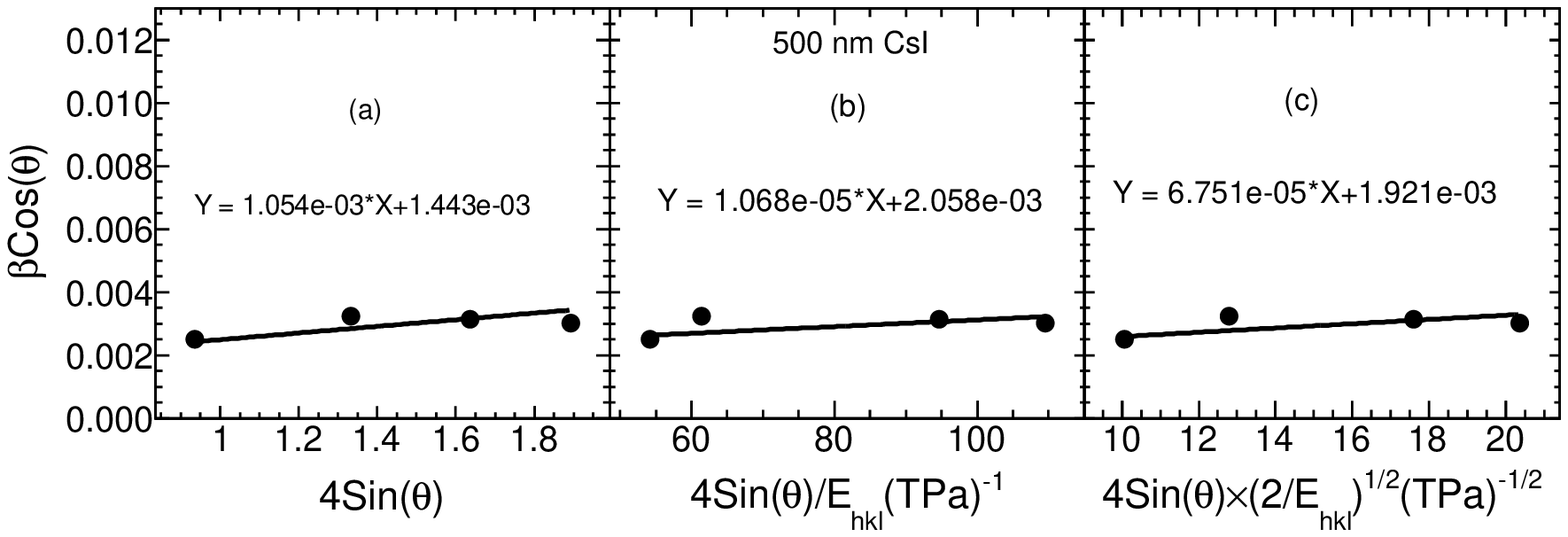}
    \includegraphics[scale=0.6]{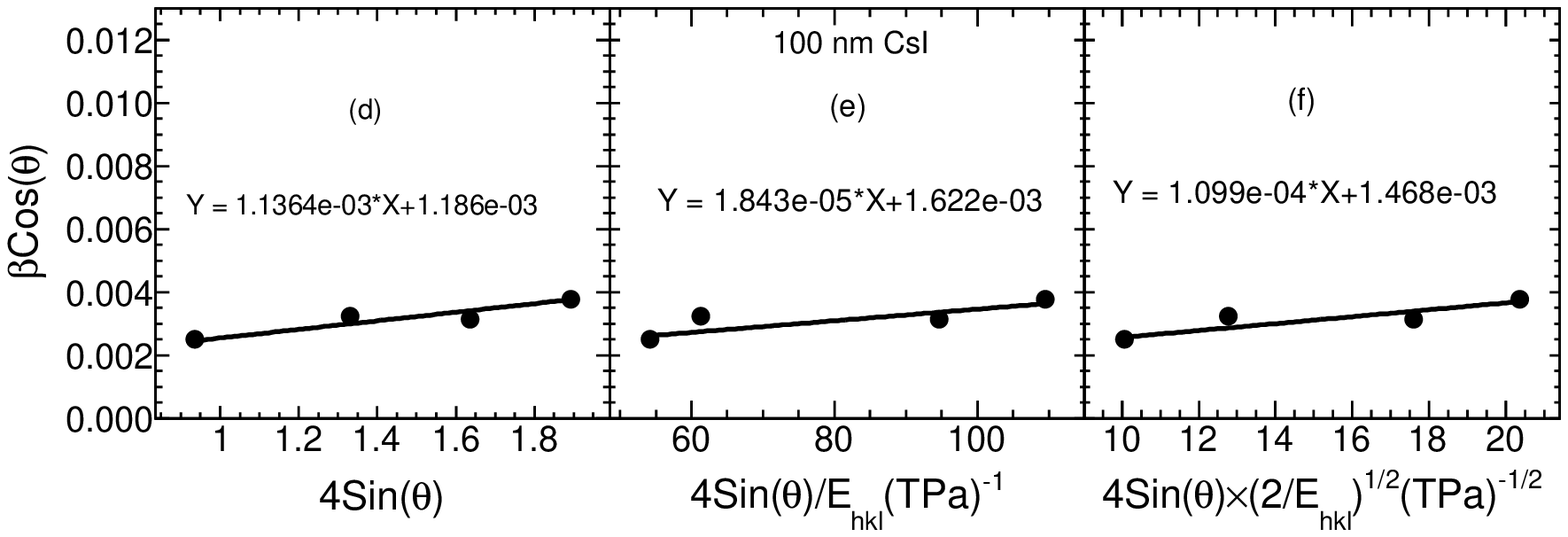}
    \caption{{Williamson-Hall plots of 500 nm and 100 nm CsI film assuming (a, d) uniform deformation model (b, e) uniform deformation stress model and (c, f) uniform deformation energy density model.}}
    \label{final_100}
  \end{center}
\end{figure*}

Further, to incorporate more realistic situation, an anisotropic approach is adopted  in uniform deformation stress model. Therefore Williamson-Hall equation is modified by an anisotropic strain $\epsilon = \sigma/E_{hkl}$, where $E_{hkl}$ is the {Young}'s modulus in direction hkl and $\sigma$ is the stress. The modified equation is written as:

\begin{equation}
  \beta_{hkl}\cos\theta = \frac{k\lambda}{D} + \frac{4\sigma\sin\theta}{E_{hkl}}
\end{equation} 

here $E_{hkl}$ for a cubic system in the direction of unit vector $l_{i}$, can be calculated using the following equation:  

\begin{eqnarray}
  \frac{1}{E_{hkl}}&=& s_{11} -2\left (s_{11} - s_{12} - \frac{1}{2}\,s_{44} \right )\Big({l_1}^2{l_2}^2\nonumber\\
  & +& {l_2}^2{l_3}^2 + {l_3}^2{l_1}^2\Big)
\end{eqnarray}

where $s_{11}$, $ s_{12}$ and $s_{44}$ are the elastic compliances of CsI. The relations which provide the connection between the elastic compliances and the stiffness $c_{ij}$ are as follows:
\begin{equation}
  s_{11} = \frac{(c_{11}+c_{12})}{(c_{11}-c_{12})(c_{11}+2c_{12})}
\end{equation}
\begin{equation}
  s_{12} = \frac{-c_{12}}{(c_{11}-c_{12})(c_{11} + 2c_{12})}
\end{equation}
\begin{equation}
  s_{44} =\frac{1}{c_{44}}
\end{equation}

where the stiffness values are $2.434{\times}10^{11}~ {dyne}/cm^{2}$, $0.636{\times}10^{11} dyne/cm^{2}$ and $0.6316{\times}10^{11}  dyne/cm^{2}$ corresponding to $c_{11}, c_{12}$ and $c_{44}$  respectively~\cite{K.Reinitz}. 

Figure 4(b) and 4(e) show the measured value of $ \beta\cos\theta$ as a function of 4$\sin\theta/E_{hkl}$ and the uniform deformation stress $\sigma$ is calculated from the slope of the line. The anisotropic lattice strain can be calculated if $E_{hkl}$ values for CsI films are known. Crystallite size can also be estimated from the intercept on  ordinate as shown in Table 2 for 100 nm and 500 nm CsI films respectively using uniform deformation stress model (UDSM).

 However, in UDEDM (equation 8), the deformation energy density(u) is  considered as a source of strain and it is assumed to be uniform in all crystallographic directions. For an elastic system that follows the Hook's law, uniform energy per unit volume(u) can be calculated from $u=(\epsilon^{2}E_{hkl})/2$. Then equation (3) can be rewritten according to the energy and strain relation.
%~\cite{khorsand}.
\begin{equation}
  \beta_{hkl}\cos\theta = \frac{K\lambda}{D} + 4\sin\theta\,\left(\frac{2u}{E_{hkl}}\right)^{1/2}
\end{equation}

\begin{table*}[!ht]
  \renewcommand{\arraystretch}{1.0}
  \caption{Geometric parameters of CsI thin films of different thickness: (b) Crystallite size from Scherrer's method, (c,d and e) W. H. Analysis and (f) Grain size from TEM counting.} 
  \vskip .3cm
  \begin{tabular}{|p{15.8cm}|} \hline
    \centering Williamson-Hall method
  \end{tabular}\\
  \begin{tabular}{|p{1.2 cm}|p{1.5cm}|p{2.5 cm}|p{3.5 cm}|p{4cm}|p{1cm}|} \hline
    
    \centering (a)~CsI Sample&\centering (b)~Scherre's method ~~D (nm)&\centering (c)~Uniform Deformation Model (UDM)&\centering (d)~Uniform Deformation Stress Model (UDSM)&\centering (e)~Uniform Deformation Energy Density Model (UDEDM) &(f)~TEM grain size (nm) \\\hline
  \end{tabular}
  \begin{tabular}{|p{1.2 cm}|p{1.5cm}|p{1cm}|p{1.09 cm}|p{0.89cm}|p{0.89cm}|p{0.82cm}|p{0.65cm}|p{0.82cm}|p{0.65cm}|p{0.65cm}|p{1.0cm}|} 
    && \centering D (nm)& \centering Strain $(\epsilon)\times10^{-4}$& \centering D (nm)& \centering Stress $(\sigma)$  MPa& \centering Strain $(\epsilon)\times10^{-4}$&\centering D (nm)&Energy Density (u) $kJ m^{-3}$&  \centering Stress $(\sigma)$  MPa&\centering Strain $(\epsilon)\times10^{-4}$&  \\\hline
    \raggedleft500 nm&\centering 54.74& \centering 95.02&\centering 10.54&\centering 66.62&\centering 10.68&\centering 5.8&\centering 71.37&\centering 4.56&\centering 12.95&\centering 7.03&306\\\hline
    \raggedleft100 nm&\centering 54.74&\centering 115.6&\centering 11.36&\centering 84.53&\centering 18.43&\centering 10.02&\centering 93.40&\centering12.08&\centering 25.68&\centering 13.96&303  \\\hline
    \raggedleft20 nm&\centering 55.0&&&&&&&&&&116  \\\hline
    \raggedleft 4 nm&\centering 41.0&&&&&&&&&&42\\\hline
  \end{tabular}
\end{table*}

Uniform deformation energy density (u) can be calculated from the slope of the line plotted between $\beta_{hkl}\cos\theta$ and $4\sin\theta(2/E_{hkl})^{1/2}$ as shown in Figure 4(c) and 4(f). The strain can also be calculated by knowing the   $E_{hkl}$ values and is reported in Table 2. The {Young}'s modulus ($E_{hkl}$) has been calculated and resulted to be 17.2873 GPa for (110) lattice plane followed by $E_{hkl}$ = 21.7048 GPa for (200), $E_{hkl}$ = 17.2873 GPa for (211) and $E_{hkl}$ = 17.2873 GPa for (220) lattice plane. Table 2 summarizes the geometrical parameters of CsI films of different thickness obtained from Debye-Scherrer's formula, various methods of W-H analysis and TEM measurements. 

The average value of crystallite size, internal strain and stress obtained from the various models of modified W-H analysis are different, thus indicating that the inclusion of strains in various form of W-H analysis have an impact on the average crystallite size of CsI films. However, there is a variation between the crystallite size obtained from Debye-Scherrer's equation and the modified W-H analysis. This difference might be due to the strain contribution to the peak broadening in thin films.

A well aligned X-ray diffractometer is used for the present study. However, errors due to finite step size of measurement in determining $2\theta$ are considered and propagated properly. The error bars are within the experimental data points in Figure 4 and the correlation coefficients in case of 4(a), 4(b) and 4(c) are 0.7, 0.5 and 0.6  while in case of 4(d), 4(e) and 4(f) they are 0.9, 0.8, and 0.8 respectively, showing a good correlation between the data points.

The results are summarized in Table 2 for strain-stress analysis. The crystallite size obtained from ScherrerÕs method using equation (1) is shown in column (b). In column (c) the crystallite size and strain are mentioned from the Uniform Deformation Model using the slope and intercept from Figure 4(a) and 4(d). In column (d) the values of crystallite size, stress and strain from Uniform Deformation Stress Model calculated from Figure 4(b) and 4(e) are shown. In column (e) crystallite size, energy density, stress and strain are reported using the fitting parameters from Figure 4(c) and 4(f). Column (f) shows the TEM results for grain size which is discussed in the next section. 

\subsection{Particle size and diffraction pattern from TEM}
TEM measurements are supposed to be a better tool for  grain size determination due to the produced image of the sample. The results obtained from TEM analysis presented in Figure 5 show that in case of 4 nm  CsI film, the layer does not appear to be continuous exhibiting a surface coverage of  $ 29\%$  only. The average grain size estimated from TEM image is about   42 nm. This is in close agreement with the results obtained from Scherrer's method (see Table 2). In case of 20 nm films, layers exhibit morphology of interconnected crystallites of discontinuous structure; the average size is about 116 nm.   Thicker CsI layers exhibit quite uniform surface morphology and larger grain size than the thinner film and having columnar shape with hexagonal structure. 100 nm and 500 nm thick CsI films have average grain size of about 300 nm as shown in Figure 6. The average grain size of a particular TEM image is estimated from the grain size distributions. The size of a particular grain is calculated by using the length of scale given by TEM system. One may observe from Figure 5 that the grain size and density of grains depend on the thickness of the film.  In case of thinner CsI films, grain size as well as grain density is small and surface morphology is discontinous with small coverage of surface area. However, with increasing thickness, both grain size as well as the density of grains increases and film surface becomes fully covered.

\begin{figure}[!ht]
  \begin{center}
    \includegraphics[scale=0.37]{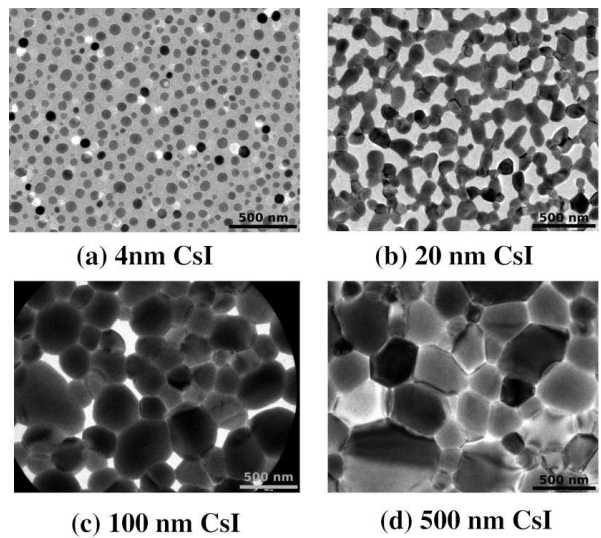}
    \caption{Transmission electron microscope (TEM) surface image of a) 4 nm, b) 20 nm, c) 100 nm and d) 500 nm Òas-depositedÓ CsI thin films.}
    \label{Sem_image}
  \end{center}
\end{figure}

Figure 7 shows selected area electron diffraction (SAED)  patterns of CsI thin film of various thicknesses i.e. (a) 4 nm, (b) 20 nm, (c) 100 nm and (d) 500 nm respectively. 

In SAED patterns, the close examination of rings reveals that they consist of a large number of spots, each arising from Bragg's reflection from an individual crystallite. Although in case of polycrystalline specimens, the diffraction spots occur at all azimuthal angles and give the appearance of continuous rings if many grains lie within the path of the electron beam (grain size $<<$ beam diameter at the specimen).
  
  \begin{figure}[!ht]
    \begin{center}
      \includegraphics[scale=0.45]{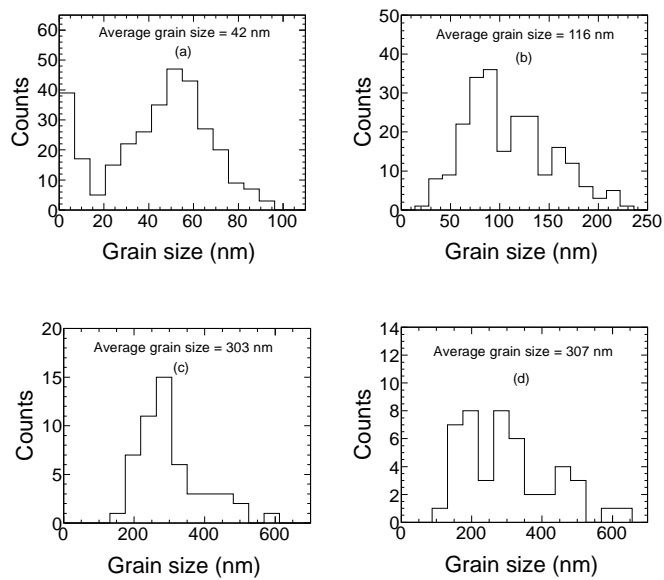}
      \caption{Grain size distribution obtained from transmission electron microscope (TEM) surface image of a) 4 nm, b) 20 nm, c) 100 nm and d) 500 nm "as-deposited" CsI thin films.}
      \label{Sem_image}
    \end{center}
\end{figure}
  
  It has been observed that the SAED pattern obtained from CsI thin films of various thicknesses are crystalline in nature. The SAED pattern of 4 nm CsI thin film demonstrates that the film has randomly oriented grains like a polycrystalline specimen. However, SAED patterns obtained for 20 nm, 100 nm and 500 nm CsI thin films show a discrete lattice of sharp spots which demonstrates that the  films have single crystal domains. The crystallographic planes obtained from CsI thin film corresponds to a body centered cubic (bcc) structure with lattice constant $a = 4.666\AA$. 
  
  \begin{figure}[!ht]
    \begin{center}
      \includegraphics[scale=0.45]{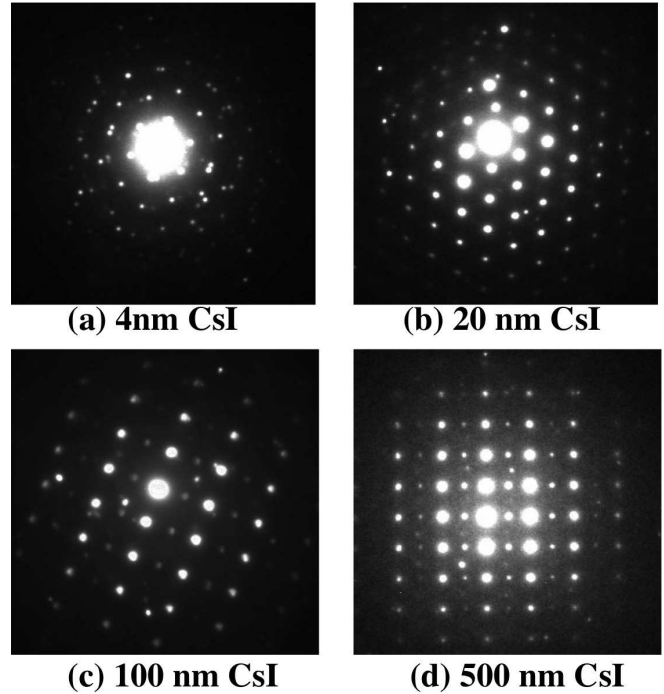}
      \caption{The electron diffraction pattern obtained from as-deposited CsI thin films of a) 4 nm, b) 20 nm, c) 100 nm and d) 500 nm thickness.}
      \label{Tem_image}
    \end{center}
  \end{figure}
     
      By comparing the results for crystallite size obtained from XRD and TEM analysis, there is a good agreement for 4 nm CsI film. However, for the sample with increasing thickness, there is an apparent difference between the grain and crystallite sizes obtained by these two methods in which grain size measured by TEM counting is higher than that the crystallite size from XRD analysis. When the thickness of film is increased from 4 nm to 500 nm, crystallite size obtained from Scherrer's equation remains constant, however in case of TEM measurements, grains size increases sharply. It indicates that according to the results from equation (1), grain growth settles to be saturated around 20 nm and further adding more thickness does not boost the crystallite size. However, W-H analysis suggest that by increasing the film thickness from 100 nm to 500 nm, the crystallite size and strain decreases. The results from TEM analysis suggest that the grain size of thicker films such as 100 nm and 500 nm is much larger than the crystallite size obtained from XRD analysis (see Table 2). The reason behind the size variation obtained from these two different techniques (XRD and TEM) can be understood in the following way: crystallite size obtained from XRD is the measurement of coherently scattering domain normal to the diffracting planes, having same orientation. While the grain size obtained from the TEM measurement is the cluster of such coherently scattering domain separated by the sharp contours (grain boundary). Further, this variation can be understood in terms of dislocations. When dislocations are arranged in a configuration causing small orientation differences between two adjacent regions, crystallite size obtained from the XRD shows two different regions. On the other hand, these two regions seem to be merged into one (single bigger grain) due to the quite small orientation difference and the contrast difference between them is not visible in TEM technique. Therefore the boundary is not considered as grain boundary in TEM technique~\cite{Tamas ungar, bolmaro}.

      \section{Conclusion}
      CsI films of various thicknesses were deposited by thermal evaporation technique and are characterized by XRD and TEM measurements. The displacement of (110) diffraction peaks towards the lower side of $\theta$ from their corresponding crystal data indicates that tensile stress exists for all CsI samples. The line broadening of as-deposited CsI films due to small crystallite size is analyzed by Debye-Scherrer's formula.
      A modified W-H method is used to estimate the crystallite size, and strain induced broadening due to the lattice deformation. Further, the origin of internal stress in a thin film comes from lattice defects such as dislocations, due to lattice misfit with it's substrate and due to differential thermal expansion between the film and it's substrate etc. In the present work, small values of stress suggest less density of lattice defects in our prepared CsI thin films.   
      
      Further, both XRD and TEM measurements show that for 4 nm thin CsI film, the grain size and crystallite size are comparable. While for other films with 20 nm, 100 nm and 500 nm thickness, TEM provides grain size larger than the crystallite size calculated with XRD analysis. It indicates that for very small grain size regime there is a good correlation between TEM and XRD but in larger grain size regime TEM counting provides a larger average grain size than crystallite size from XRD. It suggest that as we increase the thickness, the coherent domains start merging and make a bigger grain. Also by increasing the thickness from 100 nm to 500 nm, although the grain size increases, but the coherent scattering domains start decreasing. Further the difference in crystallite size from W-H analysis may be due to the variation of strain treatment within three models.
      
      To the best of our knowledge, a detailed study using UDM, UDSM and UDEDM on the CsI films is not reported yet. We may suggest that these models can be precisely used for the estimation of crystallite size and strain of CsI films.

      \section{Acknowledgment}
      This work was partially supported by the Department of science and technology (DST), the council of scientific and industrial research (CSIR) and by Indian Space Research Organization (ISRO), Govt. of India. Triloki acknowledges the financial support obtained from UGC under research fellowship scheme for meritorious students (RFSMS) program and P. Garg acknowledges the financial support from CSIR, New Delhi, India.  
      
      %\appendix
      
      
      %%\bibliographystyle{elsarticle-num} 
      %\bibliography{<your bibdatabase>}
      

\end{document}